\documentclass[aps,preprint]{revtex4}%
\usepackage{amsfonts}
\usepackage{amsmath}
\usepackage{amssymb}
\usepackage{graphicx}%
\begin{document}
\preprint{ }
\title[ ]{Structural Behavior of Non-Oxide Perovskite Superconductor MgCNi$_{3}$ at
Pressures up to 32 GPa}
\author{Ravhi S Kumar$^{a}$, A. L. Cornelius$^{a}$, Yongrong Shen$^{a}$, T. Geetha
Kumary$^{b}$, J. Janaki$^{b}$, M. C. Valsakumar$^{b}$ and \ Malcolm F.
Nicol$^{a}$}
\affiliation{$^{a}$Department of Physics, University of Nevada Las Vegas, Nevada 89154, USA}
\affiliation{$^{b}$Materials Science Division, IGCAR, Kalpakkam, India}

\begin{abstract}
We report the pressure dependence of the structural parameters of the
non-oxide perovskite superconductor MgCNi$_{3}$ up to 32 GPa using a diamond
anvil cell and synchrotron x-rays at room temperature. \ The structure of the
compound remains in the Pm-3m cubic symmetry throughout the pressure range.
The bulk modulus $B_{0}=156.9\pm0.2$GPa with $B_{0}^{\prime}=9.8$obtained by
fitting the pressure-volume data is in good agreement with theoretical
calculations reported earlier. An anomalous shift of the (111) and (200) lines
observed above 9 GPa indicates a possible local short range distortion that is
consistent with earlier studies.

\end{abstract}
\volumeyear{year}
\volumenumber{number}
\issuenumber{number}
\eid{identifier}
\date[Date text]{date}
\startpage{1}
\endpage{6}
\maketitle

\section{Introduction}

The recent discovery of superconductivity in the Ni rich perovskite
MgCNi$_{3}$ with a transition temperature $T_{c}=8.5$ K has triggered intense
research towards the search for superconducting materials in the intermetallic
family.\cite{He01} The observation of superconductivity in this compound is
unusual and surprising as Ni has strong magnetic behavior due to partially
filled \textit{d }states. The domination of Ni \textit{3d} bands inferred from
band structure calculations emphasizes a strong hybridization between the Ni
\textit{3d} and C \textit{2p} electrons,\cite{Sing01,Shim01} and the
non-ferromagnetic ground state of MgCNi$_{3}$ is consequence of a reduced
Stoner factor due to this hybridization. Neutron diffraction experiments
performed at low temperatures down to 2 K show no sign of \ ferromagnetic
order indicating the absence of magnetic anomalies around the transition
temperature.\cite{Huang01} Ignatov \textit{et al}., performed Ni $K$-edge
x-ray absorption measurements and reported distortions in the Ni$_{6}$
octahedra below 70 K which favour a low symmetry crystal
structure.\cite{Ignatov03} The ferromagnetic spin fluctuations observed in the
NMR experiments,\cite{Singer01} and the unusual quasi two dimensional van-Hove
singularity (vHs) reported by Rosner\textit{ et al.},\cite{Rosner02} suggest
that MgCNi$_{3}$ lies in the proximity of a ferromagnetic boundary. There are
still open questions regarding the role of spin fluctuations and/or lattice
instabilities on the origin of superconductivity in this system.

Even though it is reported that the density of states shows normal behavior
under pressure up to 20 GPa, $T_{c}$ shows an increase as a function of
pressure.\cite{Kumary02,Yang03,Garbarino04} The possible reason behind the
rise in $T_{c}$ has been attributed to either a reduction in the spin
fluctuations or an increase in the electron-phonon interaction followed by a
structural transition. The recent temperature dependent \ inelastic neutron
scattering studies give evidence for a lattice instability in the low
frequency Ni phonon modes. Still, there have been no detailed structural
studies on this system to clarify the structural stability at high pressure,
with the exception of an energy dispersive x-ray diffraction (EDXRD) report by
Youlin \textit{et al}.\cite{Youlin03} In order to investigate the structural
properties under pressure, we have performed high pressure x-ray diffraction
experiments on this compound and discuss the results in detail in the
following sections.

\section{Experimental}

The sample used in the experiments was synthesized by the conventional solid
state reaction method reported earlier.\cite{Kumary02} The x-ray diffraction
patterns recorded at ambient conditions showed the compound \ crystallizes in
the cubic perovskite phase with a minor impurity of unreacted Ni (2 -5 \%).
The cell parameter obtained at ambient conditions, $a=3.8100(\pm0.0004)$
\AA ,\ matches well with the reported values in the literature.\cite{He01} The
AC susceptibility and four probe resistivity measurements showed the $T_{c}$
onset around 8 K.

High pressure experiments were performed using a Merrill-Bassett type diamond
anvil cell (DAC) with a culet diameter of 400 $\mu$m at Sector 16 IDB, HPCAT,
Advanced Photon Source (APS), Chicago. The sample in powder form was loaded
with tiny ruby chips in a 185 $\mu$m\ hole drilled in a stainless steel gasket
with a preindentation to 65 $\mu$m . We performed three experimental runs up
to 32 GPa: two with silicone fluid pressure transmitting medium and another
with Flourinet (FC70) to examine effects of pressure medium on the results.
The diffraction images were recorded using an imaging plate. In all the
experimental runs the typical beam size was 20x20 $\mu$m$^{2}$ , and the
exposure time for each pattern was 10-20 sec. The pressure in the DAC has been
determined using the standard ruby fluorescence method. The diffraction images
were integrated using the Fit2D software\cite{Hammersley94} and the structural
refinement has been done using RIETICA (LHPM) Rietveld package\cite{Howard98}
and JADE.\cite{jade}

\section{Results and discussion}

The crystal structure of MgCNi$_{3}$ is a three dimensional network with the
Ni atoms crystallographically located at 3\textit{c} (0;1/2;1/2), Mg at
1\textit{a} (0;0;0) and C at 1\textit{b} (1/2;1/2;1/2) adopting the Pm-3m
cubic space group symmetry. The Ni atoms occupy the position of the negative
halide atoms in the common perovskite structure, and form a metallic Ni$_{6}$
octahedra frame work. The superconducting properties are strongly dependent on
the atomic position of Ni, which in turn governs the Ni-C bonding and Ni-Ni
hopping.%
%TCIMACRO{\FRAME{ftbpFU}{3.5751in}{2.7043in}{0pt}{\Qcb{X-ray diffraction
%patterns collected at different pressures (Run 1). The star symbol denotes the
%diffraction lines due to unreacted Ni.}}{\Qlb{fig1}}{kumar1.eps}%
%width 3.5751in;  height 2.7043in;  depth 0pt;  original-width 3.5284in;
%original-height 2.661in;  cropleft "0";  croptop "1";  cropright "1";
%filename '../../../sw50/docs/kumar1.eps';file-properties "XNPEU";}}}%
%BeginExpansion
\begin{figure}
[ptb]
\begin{center}
\includegraphics[
height=2.7043in,
width=3.5751in
]%
{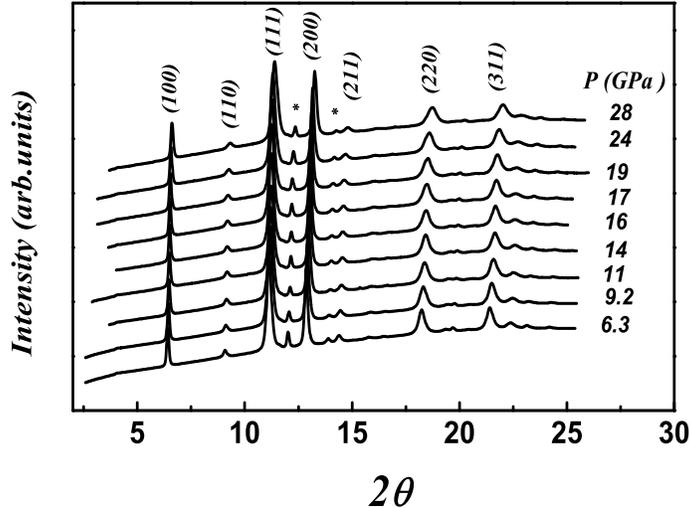}%
\caption{X-ray diffraction patterns collected at different pressures (Run 1).
The star symbol denotes the diffraction lines due to unreacted Ni.}%
\label{fig1}%
\end{center}
\end{figure}
%EndExpansion

X-ray diffraction patterns collected at several pressures are shown in the
Fig. 1. In general the diffraction lines observed can be clearly indexed to
the cubic structure. When analyzing the pressure data after the first run and
refining the structural parameters, we noticed a strong deviation from the
systematic shift of the (111) and (200) lines under pressure above 8 GPa (Fig.
2).
%TCIMACRO{\FRAME{ftbpFU}{3.8787in}{3.5803in}{0pt}{\Qcb{Rietveld refinement for
%the diffraction pattern collected for MgCNi$_{3}$ at 0.6 GPa in Run 1. The
%residuals are $R_{wp}$=1.2 \% and $\chi^{2}$ = 1.1. \ The upper panel shows
%the diffraction lines (111) and (200) after refinement using cubic symmetry in
%Run 3 at 10.6 GPa. The solid line represents the calculated spectrum and open
%symbols represent the observed data.}}{\Qlb{fig2}}{kumar2.eps}%
%width 3.8787in;  height 3.5803in;  depth 0pt;  original-width 3.8303in;
%original-height 3.5336in;  cropleft "0";  croptop "1";  cropright "1";
%filename '../../../sw50/docs/kumar2.eps';file-properties "XNPEU";}}}%
%BeginExpansion
\begin{figure}
[ptb]
\begin{center}
\includegraphics[
height=3.5803in,
width=3.8787in
]%
{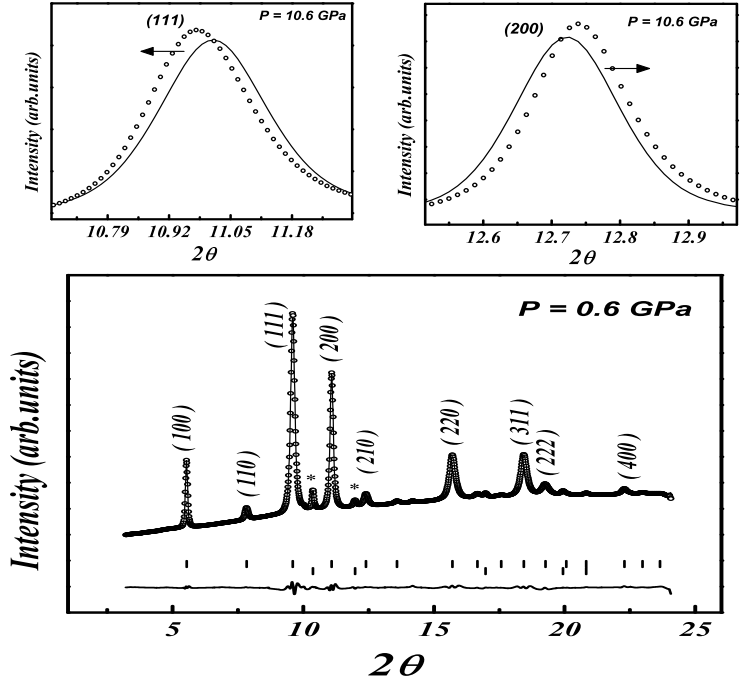}%
\caption{Rietveld refinement for the diffraction pattern collected for
MgCNi$_{3}$ at 0.6 GPa in Run 1. The residuals are $R_{wp}$=1.2 \% and
$\chi^{2}$ = 1.1. \ The upper panel shows the diffraction lines (111) and
(200) after refinement using cubic symmetry in Run 3 at 10.6 GPa. The solid
line represents the calculated spectrum and open symbols represent the
observed data.}%
\label{fig2}%
\end{center}
\end{figure}
%EndExpansion
The (111) line tends to stay at low angles and (200) shifts rapidly to higher
angles with increasing pressures. As the cubic perovskites are very sensitive
to the uniaxial stress distributions around the sample environment, usually
one expects uneven shifts and splittings, even pressure induced changes to
rhombohedral symmetry if a non-hydrostatic stress develops during the glassy
transformation of the pressure transmitting medium.\cite{Takemura01,Singh04}
Since our recent high pressure experiments found the silicone fluid medium
(poly dimethyl siloxane with a viscosity of 1 cst) to be nearly hydrostatic at
least up to 10 GPa,\cite{Shen04} the effect of non hydrostatic stress on the
sample in the first run is expected to be less pronounced below 10 GPa. To see
whether these anomalies are due to the pressure medium or intrinsic local
distortions associated with structural modifications, we decided to repeat the
experiments (Run 2 with silicone fluid and Run 3 with Fluorinert - FC70). The
hydrostatic limit of Fluorinert (FC70) has been recently reported to be 0.55
GPa.\cite{Varga03} Thus when using Fluorinert pressure medium, one might
expect the anomalous line shifts to appear at a lower pressure relative to
silicone fluid if the cause of the shifts is due to the solidification of the
pressure medium. The results of the refinements of the data obtained from Run
2 and 3 showed similar behavior observed in Run 1 around 8 GPa which showed
that the stress distributions due to change in the pressure medium are not
significant. This indicates that there are intrinsic local distortions from
the ideal cubic structure.

MgCNi$_{3}$ is isostructural to the well known Bi$_{1-x}$K$_{x}$BiO$_{3}$
(BKBO) superconductor which has $T_{c}$ = 30K. Structural distortions in BKBO
have been studied in detail by Braden \textit{et al.,} and they have shown
that the rotational instability of the BiO$_{6}$ octahedra leads to a
tetragonal distortion.\cite{Braden00} A non-cubic layered structure for BKBO
is further reported by Klinkova \textit{et al.}\cite{Klinkova03} As
MgCNi$_{3}$ and BKBO both have breathing instabilities and structural
similarities, we carefully examined the high pressure x-ray diffraction
patterns of MgCNi$_{3}$ for super cell reflections and splits in diffraction
lines for possible structural transitions. The spectra showed no considerable
line broadening and no super cell reflections or splittings up to the highest
pressure achieved in this experiment. These observations lead us to conclude
that the distortions are associated with a change in the short range
structural order. The Rietveld refinement is confined to pressures less than 9
GPa for cubic symmetry and at higher pressures we have obtained the cell
parameters by fitting the peak positions using JADE. \ The $P-V$ data are
fitted to the Birch-Murnaghan equation of state given by\bigskip%

\begin{equation}
P=\frac{3}{2}B_{0}\left[  \left(  \frac{V}{V_{0}}\right)  ^{-7/3}-\left(
\frac{V}{V_{0}}\right)  ^{-5/3}\right]  \times\left\{  1+\frac{3}{4}\left(
B_{0}^{\prime}-4\right)  \left[  \left(  \frac{V}{V_{0}}\right)
^{-2/3}-1\right]  \right\}  .
\end{equation}
This fit yields a bulk modulus of \ $B_{0}=156.9\pm0.2$ GPa \ with a pressure
derivative $B_{0}^{\prime}=9.8$. The $P-V$ data are shown in Fig. 3.
%TCIMACRO{\FRAME{ftbpFU}{4.2557in}{3.5526in}{0pt}{\Qcb{$P-V$ data for
%MgCNi$_{3}$. The open symbols represent EDXRD data from Ref. 11. }}%
%{\Qlb{fig3}}{kumar3.eps}{\special{ language "Scientific Word";
%type "GRAPHIC";  maintain-aspect-ratio TRUE;  display "USEDEF";
%valid_file "F";  width 4.2557in;  height 3.5526in;  depth 0pt;
%original-width 4.2056in;  original-height 3.506in;  cropleft "0";
%filename '../../../sw50/docs/kumar3.eps';file-properties "XNPEU";}}}%
%BeginExpansion
\begin{figure}
[ptb]
\begin{center}
\includegraphics[
height=3.5526in,
width=4.2557in
]%
{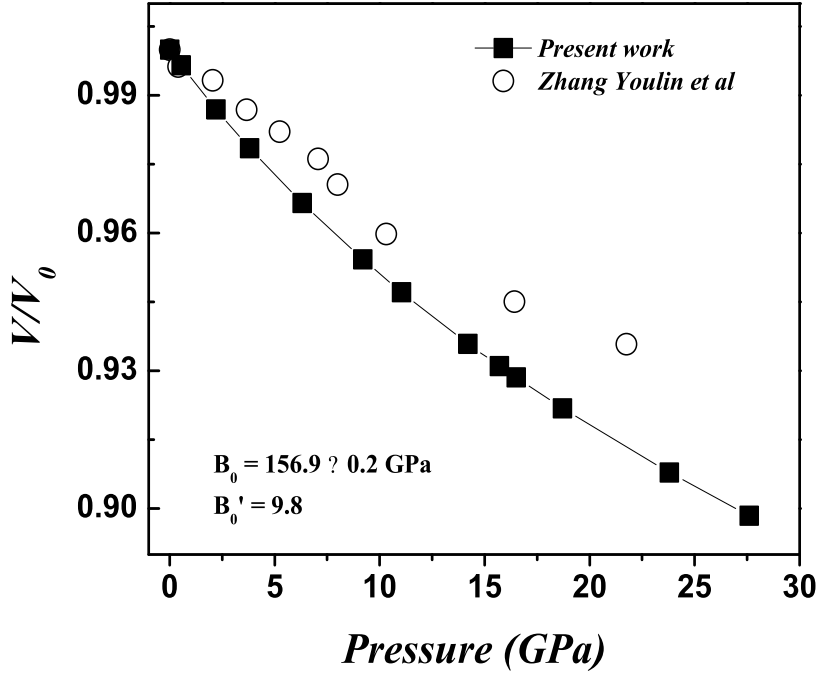}%
\caption{$P-V$ data for MgCNi$_{3}$. The open symbols represent EDXRD data
from Ref. 11. }%
\label{fig3}%
\end{center}
\end{figure}
%EndExpansion
The bulk modulus obtained in our experiment agrees well with the theoretical
value estimated earlier\cite{Kumary02} and is a factor of 1.7 smaller than the
reported value in the EDXRD experiment.\cite{Youlin03} The possible reason for
the high value of the bulk modulus in the EDXRD experiment may be due to the
inaccuracy in estimating the ambient\ unit cell volume, pressure determination
using the internal calibrant and/or the distortions induced on the unit cell
due to non-hydrostatic stress coupled with local distortions as there is no
use of pressure medium reported. The bulk modulus of MgCNi$_{3}$ is comparable
to MgB$_{2}$, high $T_{c}$ superconducting compounds and other oxide
perovskites and it is more compressible than its two dimensional borocarbide
analogues such as YNi$_{2}$B$_{2}$C$_{2}$ [11,
21-26].\cite{Youlin03,Vogt01,Meenakshi98,Rossler98,Kaye95Kaye95,Akhtar94,Pruzan02}
Our experimental results reflect the fact that the Ni-Ni and Ni-C bonding in
MgCNi$_{3}$ are softer than ternary borocarbides and the Bi-O bonds in
bismuthates. We have listed the bulk moduli value of some perovskites with
elemental Ni in Table 1 for comparison.\begin{table}[ptbptb]
\narrowtext
\par%
\begin{tabular}
[c]{|c|c|c|c|c|c|}\hline
Compound & Structure & $T_{c}$ (K) & $B_{0}$ (GPa) & $B_{0}^{\prime}$ &
Ref.\\\hline
MgCNi$_{3}$ & Cubic, Pm-3m & 8 & 156 & 9.9 & this work\\\hline
MgCNi$_{3}$ & Cubic, Pm-3m & 8 & 267 & 4 & 11\\\hline
MgB$_{2}$ & Hexaganol, P6/mmm & 40 & 151 & 5 & 21\\\hline
YNi$_{2}$B$_{2}$C & Tetraganol, I4/mmm & 15 & 200 & - & 22\\\hline
TbNi$_{2}$B$_{2}$C & Tetraganol, I4/mmm & 0.3 & 136-196 & - & 23\\\hline
Ni & Cubic, Fm-3m & - & 177-187 & - & 24\\\hline
Ba$_{0.6}$K$_{0.4}$BiO$_{3}$ & Cubic, Pm-3m & 30 & 200 & 4 & 25\\\hline
BaBiO$3$ & Tetragonal, P-42 m & - & 215 & 4 & 25\\\hline
BaTiO$_{3}$ & Cubic, Pm-3m & - & 135 & 6.4 & 26\\\hline
KNbO$_{3}$ & Cubic, Pm-3m & - & 146 & 5 & 26\\\hline
\end{tabular}
\caption{Bulk modulus data and other physical properties of some perovskites
and borocarbides.}%
\label{table}%
\end{table}

In conclusion we have performed high pressure x-ray diffraction measurements
on the non-oxide perovskite MgCNi$_{3}$ up to 32 GPa. The determined bulk
modulus is in excellent agreement with the theoretical value predicted earlier
by TB-LMTO calculations. Even though the results show no structural phase
transitions, we have noticed a pressure induced distortion which may be
associated with a change in the short range order of the crystal structure.
Detailed extended x-ray absorption (EXAFS) measurements under pressure are
under way to understand the pressure induced changes in this system.

\begin{acknowledgments}
The authors thank Drs. Maddury Somayazulu and Liu from HPCAT for their
technical help. Work at UNLV is supported by DOE EPSCoR-State/National
Laboratory Partnership Award DE-FG02-00ER45835. HPCAT is a collaboration among
the UNLV High Pressure Science and Engineering Center, the Lawrence Livermore
National Laboratory, the Geophysical Laboratory of the Carnegie Institution of
Washington, and the University of Hawaii at Manoa. The UNLV High Pressure
Science and Engineering Center was supported by the U.S. Department of Energy,
National Nuclear Security Administration, under Cooperative Agreement
DE-FC08-01NV14049. Use of the Advanced Photon Source was supported by the U.
S. Department of Energy, Office of Science, Office of Basic Energy Sciences,
under Contract No. W-31-109-Eng-38.
\end{acknowledgments}

{\Large
\bibliographystyle{apsrev}
\bibliography{acompat,MGCNI3}
}

\end{document}